\documentclass[onecolumn,superscriptaddress,nofootinbib,notitlepage,longbibliography,aps]{revtex4-1}
\pdfoutput=1

\usepackage{graphicx}
\usepackage{amsmath}
\usepackage{amssymb}
\usepackage{amsfonts}
\usepackage[dvipsnames]{xcolor}
\usepackage[linktoc=none]{hyperref}
\usepackage[utf8]{inputenc}

\hypersetup{colorlinks=true,linkcolor=blue,citecolor=teal,filecolor=magenta, urlcolor=cyan}

\newcommand{\INFN}{INFN - Sezione di Napoli, Complesso Univ. Monte S. Angelo, I-80126 Napoli, Italy}
\newcommand{\UNINA}{Dipartimento di Fisica "Ettore Pancini", Universit\'a degli studi di Napoli "Federico II", Complesso Univ. Monte S. Angelo, I-80126 Napoli, Italy}
\newcommand{\SSM}{Scuola Superiore Meridionale, Universit\'a degli studi di Napoli "Federico II", Largo San Marcellino 10, 80138 Napoli, Italy}

\begin{document}

\title{Heavy decaying dark matter at future neutrino radio telescopes}

\author{Marco Chianese}
\email{marco.chianese@unina.it}
\affiliation{\UNINA}
\author{Damiano F.G. Fiorillo}
\email{dfgfiorillo@na.infn.it}
\affiliation{\UNINA}
\affiliation{\INFN}
\author{Rasmi Hajjar}
\email{rasmienrique.hajjarmunoz@unina.it}
\affiliation{\SSM}
\author{Gennaro Miele}
\email{miele@na.infn.it}
\affiliation{\UNINA}
\affiliation{\INFN}
\affiliation{\SSM}
\author{Stefano Morisi}
\email{smorisi@na.infn.it}
\affiliation{\UNINA}
\affiliation{\INFN}
\author{Ninetta Saviano}
\email{nsaviano@na.infn.it}
\affiliation{\INFN}
\affiliation{\SSM}

\date{\today}
\begin{abstract}
In the next decades, ultra-high-energy neutrinos in the EeV energy range will be potentially detected by next-generation neutrino telescopes. Although their primary goals are to observe cosmogenic neutrinos and to gain insight into extreme astrophysical environments, they can also indirectly probe the nature of dark matter. In this paper, we study the projected sensitivity of up-coming neutrino radio telescopes, such as RNO-G, GRAND and IceCube-gen2 radio array, to decaying dark matter scenarios. We investigate different dark matter decaying channels and masses, from $10^7$ to $10^{15}$ GeV. By assuming the observation of cosmogenic or newborn pulsar neutrinos, we forecast conservative constraints on the lifetime of heavy dark matter particles. We find that these limits are competitive with and highly complementary to previous multi-messenger analyses.
\end{abstract}

\maketitle

\section{Introduction \label{sec:intro}}

The existence of Dark Matter (DM) is one of the pillars of the standard cosmological model, yet it has escaped direct observation \textit{via} non-gravitational interactions until now. Among the possible ways to probe DM interactions beyond the gravitational one, a promising method is the search for indirect signatures of DM decay or annihilation. These processes can indeed lead to the production of an astrophysical signal of cosmic rays, gamma rays and neutrinos. In the last decade, the dawning of high-energy multi-messenger astronomy has led to powerful improvements in the indirect searches for dark matter. A key role has been played by experiments devoted to the observations of gamma rays, \textit{e.g.} Fermi-LAT~\cite{Ackermann:2014usa, Pieri:2009je}, KASKADE~\cite{Apel:2017ocm}, CASA-MIA~\cite{PhysRevLett.79.1805}, CASA-BLANCA~\cite{Cassidy:1997sb}, and of cosmic rays, \textit{e.g.} PAMELA~\cite{Picozza:2006nm}, AMS-02~\cite{Aguilar:2016kjl, Aguilar:2019owu}, TA~\cite{Kawai:2008zza}, KASKADE-Grande~\cite{Kang:2019tgo} and the Pierre Auger Observatory (PAO)~\cite{Abraham_2010}. The measurement of astrophysical neutrinos made by IceCube~\cite{Aartsen:2013jdh,Abbasi:2020jmh} has allowed to complement these searches with a study of astrophysical neutrinos in the energy range below 10 PeV. Compared to gamma and cosmic rays, neutrinos are weakly interacting elusive particles, which have the advantage of being undeflected by the galactic and extragalactic magnetic fields and unattenuated by collisions with any kind of foregrounds. The collected information from all the three messengers has allowed us to probe the decay rate and the annihilation cross section of heavy dark matter particles with masses larger than $100~\mathrm{TeV}$~\cite{Murase:2012xs,Esmaili:2012us,Feldstein:2013kka,Esmaili:2013gha,Rott:2014kfa,Boucenna:2015tra,Esmaili:2015xpa,Cohen:2016uyg,Aartsen:2018mxl,Kachelriess:2018rty,Xu:2018kwo,Bhattacharya:2019ucd,Chianese:2019kyl,Dekker:2019gpe,Ishiwata:2019aet,Arguelles:2019boy,Arguelles:2019ouk,Kalashev:2020hqc}.

In this framework, the energy range above 10 PeV is still essentially unexplored from the viewpoint of neutrino observations. Up until now, the most important information in this range comes from the PAO~\cite{Aab:2017njo} and ANITA~\cite{Gorham:2019guw}, which have placed only upper limits on the neutrino flux. A step further in the investigation of ultra-high-energy (UHE) neutrinos will be constituted by the next-generation neutrino radio telescopes which are under construction or in a design phase. These include several planned experiments, \textit{e.g.} ARIANNA~\cite{Anker:2020lre}, ARA~\cite{Miller_2012}, TAMBO~\cite{Wissel:2019alx, Romero-Wolf:2020pzh}, POEMMA~\cite{Olinto:2020oky}, JEM-EUSO~\cite{Takizawa:2007ar}, RNO-G~\cite{Aguilar:2020xnc}, GRAND~\cite{Alvarez-Muniz:2018bhp} and IceCube-Gen2 radio array~\cite{Aartsen:2019swn, Aartsen:2020fgd}. Their primary aim is the determination of the neutrino sources in this ultra-high energy range, and in particular the detection of the so-called cosmogenic neutrinos~\cite{Beresinsky:1969qj}. These neutrinos represent a guaranteed component that originates from the collision of cosmic rays on the Cosmic Microwave Background (CMB). The magnitude of this flux is however subject to severe theoretical uncertainties, connected with the chemical composition of the UHE cosmic rays. Furthermore, a competitive source of neutrinos at these ultra-high energies may be constituted by hadronic production in astrophysical environments such as active galactic nuclei~\cite{Murase:2014foa,Rodrigues:2020pli}, gamma-ray bursts~\cite{Murase:2007yt}, flat-spectrum radio quasars~\cite{Righi:2020ufi}, black-hole jets embedded in large-scale structures~\cite{Fang:2017zjf} and newborn pulsars~\cite{Fang:2013vla}. The upcoming radio telescopes could allow us to shed light on the question of what is the dominant source of astrophysical neutrinos in this energy range. At the same time, these experiments have the potential of probing a still unexplored parameter space of heavy dark matter particles with masses larger than 100 PeV. Together with gamma-ray and cosmic-ray observations, UHE neutrinos will provide a complementary and potentially competitive way of investigating dark matter indirect signals.

In this paper, we study the parameter space of decaying dark matter that could be probed and constrained with future observations of planned neutrino radio telescopes. We focus on the case of decaying DM particles with masses between $10^7$~GeV and $10^{15}$~GeV. For this mass range, annihilating DM particles are in general expected to produce only much smaller neutrino fluxes according to the unitarity bound~\cite{Griest:1989wd,Smirnov:2019ngs,Arguelles:2019ouk}. We limit this study to four benchmark neutrino radio telescopes, namely RNO-G, GRAND10k, GRAND200k and IceCube-Gen2 radio array. Taking into account the different sensitivities of these experiments, we first estimate the detection prospects to DM neutrino signals, which are assumed to dominate the possible astrophysical emissions. Then, we perform a forecast analysis to determine the constraints that can be placed on the lifetime of decaying DM particles with masses between $10^7$~GeV and $10^{15}$~GeV. In this case, we consider two different benchmark scenarios where the telescopes will observe UHE neutrinos with a cosmogenic origin and produced by newborn pulsars. Within the theoretical uncertainties of the two astrophysical scenarios, we consider the highest prediction for the neutrino flux in order to provide robust and conservative constraints.

The paper is organized as follows. In section~\ref{sec:NeuFlux}, we describe the neutrino production from dark matter decays in the galactic halo and in intergalactic space. In section~\ref{sec:neutel}, we briefly review the expected properties of the next-generation radio telescopes which are considered in this study. In section~\ref{sec:detect}, we discuss the detection prospects of DM neutrino signals, while in section~\ref{sec:ConsCons} we forecast conservative constraints to DM properties. Finally, in section~\ref{sec:conclusions} we draw the main conclusions of this work.

\section{Neutrino fluxes from heavy dark matter}
\label{sec:NeuFlux}

Dark matter particles accumulated in the galactic halo and distributed in the intergalactic space can give rise to a flux of UHE neutrinos through their decay and annihilation into ordinary matter. Our emphasis on indirect searches of decaying, rather than annihilating, dark matter is connected with the unitarity bound on DM cross section~\cite{Griest:1989wd, Smirnov:2019ngs}, which leads to much higher expected fluxes in the former rather than in the latter case. In fact, by an order-of-magnitude estimate similar to Ref.~\cite{Feldstein:2013kka}, the dominant galactic neutrino flux from decaying dark matter is of the order of
\begin{equation}
    \phi^\text{dec}_\nu\sim \frac{\rho_\text{DM}}{m_\text{DM}}\frac{1}{\tau_\text{DM}} L \sim 10^{-16}~{\rm cm^{-2} s^{-1}}\left(\frac{\rho_\text{DM}}{0.4~{\rm GeV/cm^3}}\right)\left(\frac{10^9~{\rm GeV}}{m_\text{DM}}\right)\left(\frac{10^{29}~{\rm s}}{\tau_\text{DM}}\right)  \,,
\end{equation}
where $\rho_\text{DM} \simeq 0.4~\rm GeV/cm^3$ is the typical galactic density of DM particles, $\tau_\text{DM}$ is their lifetime, $m_\text{DM}$ is the DM mass, and $L \simeq 10~\rm kpc$ is the length scale of our galaxy (here and throughout this work we adopt natural units in which $c=\hbar=1$). In particular, decaying DM particles with $\tau_\text{DM}=10^{29}~\text{s}$ and $m_\text{DM} = 10^9~\text{GeV}$ are expected to produce a neutrino flux within the reach of upcoming neutrino telescopes. On the other hand, for annihilating dark matter the corresponding flux is of the order of
\begin{equation}
    \phi^\text{ann}_\nu \sim \left(\frac{\rho_\text{DM}}{m_\text{DM}}\right)^2 \sigma v_\text{DM} L,
\end{equation}
where $\sigma$ is the annihilation cross section and $v_\text{DM} \simeq 10^{-3}\,c$ is the typical velocity of dark matter particles. For $s$-wave annihilation, the cross section $\sigma$ is bounded by the unitarity value $\sigma<4\pi/(m_\text{DM} v_\text{DM})^2$. By substitution we find
\begin{equation}
 \phi^\text{ann}_\nu \lesssim 10^{-27}~{\rm cm^{-2} s^{-1}}\left(\frac{\rho_\text{DM}}{0.4~{\rm GeV/cm^3}}\right)^2 \left(\frac{10^9~{\rm GeV}}{m_\text{DM}}\right)^4 \left(\frac{10^{-3}\,c}{v_\text{DM}}\right) \,,
\end{equation}
which is far beyond the sensitivity of next-generation neutrino radio telescopes. Hence, for annihilating dark matter the number of neutrino events expected in these observatories is negligibly small. Except for peculiar scenarios featuring very dense and/or very cold dark matter substructures~(see {\it e.g.} Ref.~\cite{Zavala:2014dla}), it appears that for DM masses larger than $10^7~\text{GeV}$ neutrino observations are not able to probe the allowed parameter space of annihilating DM particles (see also Ref.~\cite{Arguelles:2019ouk}). For this reason, hereafter we will focus on the scenario of decaying dark matter only.

The neutrino flux expected from dark matter consists of a galactic and extragalactic component. The former is described by the following differential flux of neutrinos and anti-neutrinos per flavour $\alpha$:
\begin{equation}\label{eq:galDM}
    \frac{\mathrm{d}\Phi^\text{gal.}_{\nu_{\alpha}+\bar{\nu}_{\alpha}}}{\mathrm{d}E_\nu \mathrm{d}\Omega}=\frac{1}{4\pi m_\text{DM} \tau_\text{DM}}\frac{\mathrm{d}N_\alpha}{\mathrm{d}E_\nu} \int_0^\infty \mathrm{d}s\,\rho_\text{DM}[r(s,\ell,b)] \,.
\end{equation}
The quantity $\rho_\text{DM}(r)$ is the dark matter galactic distribution, which is assumed to be the commonly-adopted Navarro-Frank-White (NFW) profile~\cite{Navarro:1995iw}. It is a function of the galactic coordinates $\ell$ and $b$ and the line-of-sight distance $s$ from the Earth, which define the galactocentric radial coordinate $r=\sqrt{s^2+R_\odot^2-2sR_\odot\cos\ell \cos b}$ with $R_\odot=8.5$ kpc. The NFW profile is given by
\begin{equation}
    \rho_\mathrm{DM}(r)=\frac{\rho_s}{r/r_s(1+r/r_s)^2}\,,
\end{equation}
with $r_s=24$ kpc and $\rho_s=0.18$ GeV\,cm$^{-3}$ for the Milky Way~\cite{Cirelli:2010xx}. The diffuse neutrino flux in Eq.~\eqref{eq:galDM} depends on the energy spectrum of the $\alpha$-flavour neutrinos produced in the decay of a single DM particle $\mathrm{d}N_\alpha/\mathrm{d}E_\nu$. The computation of these neutrino spectra from DM involves a non-trivial analysis of the electroweak cascades which develop after the primary decay: for this work we use the results of the recent code \texttt{HDMSpectra} \cite{Bauer:2020jay}.\footnote{It is worth noticing that the \texttt{HDMSpectra} tabulated results are less-reliable for $x=2E_\nu/m_\mathrm{DM}<10^{-4}$~\cite{Bauer:2020jay}. This might affect the constraints we report in this analysis. However, for most of the DM masses we consider, the large contribution to the neutrino flux within reach of upcoming neutrino telescopes comes from the $x>10^{-4}$ range. Being the neutrino radio telescopes sensitive up to energies $\sim 10^{11}~\text{GeV}$, this implies that only the mass range $m_\mathrm{DM} \gtrsim \mathcal{O}(10^{14}~\text{GeV})$ is affected by the large theoretical uncertainties on the decay particle spectra.}

The extragalactic component can be similarly expressed as
\begin{equation}\label{eq:extragalDM}
    \frac{\mathrm{d}\Phi^\text{ext.gal.}_{\nu_{\alpha}+\bar{\nu}_{\alpha}}}{\mathrm{d}E_\nu \mathrm{d}\Omega}=\frac{\Omega_\text{DM} \rho_c}{4\pi m_\text{DM} \tau_\text{DM}} \int_0^\infty \frac{\mathrm{d} z}{H(z)} \frac{\mathrm{d}N_\alpha}{\mathrm{d}E^\prime_\nu}\bigg{\vert}_{E^\prime_\nu=E_\nu(1+z)} \,,
\end{equation}
where $\rho_c$ is the critical density of the Universe, $\Omega_\text{DM}$ encodes the density of the DM particles in the Universe, $z$ is the cosmological redshift and $H(z)$ is the Hubble expansion rate taking into account the latest Planck measurements of the cosmological parameters~\cite{Aghanim:2018eyx}. Differently from gamma rays, the absorption of neutrinos in the intergalactic medium is negligible.

Neutrinos will travel large distances before arriving to the detectors on the Earth, and on these large scales they oscillate very rapidly compared to the propagation length. Their flavour composition at the Earth will then be the result of very fast averaged oscillations. Since the experiments under consideration provide only an all-flavour sensitivity, we rearrange the flux in Eq.~(\ref{eq:galDM}) and Eq.~(\ref{eq:extragalDM}) into an equivalent flux that is summed over all flavours. Moreover, we are mainly concerned with angle-averaged fluxes since most of the planned experiments do not provide the sensitivity as a function of the neutrino arrival direction, but only the angle-averaged sensitivity. It is worth noticing that this does not influence our result for the isotropic extragalactic component. On the other hand, for the galactic component featuring an anisotropy towards the galactic center, the computation based on the angle-averaged flux makes our results very conservative, because it tests only the flux normalization and not its angular distribution. In addition, our analysis is not affected by the choice of the dark matter halo profile. In order to preserve the consistency of the analysis, we therefore consider the angle-integrated dark matter flux for the all-flavour neutrino case which reads as follows
\begin{equation}\label{eq:DMneuflux}
    \frac{\mathrm{d}\Phi^\text{DM}_{3\nu}}{\mathrm{d}E_\nu}= \sum_\alpha \int \mathrm{d}\Omega  \left[\frac{\mathrm{d}\Phi^\text{gal.}_{\nu_{\alpha}+\bar{\nu}_{\alpha}}}{\mathrm{d}E_\nu \mathrm{d}\Omega}+ \frac{\mathrm{d}\Phi^\text{ext.gal.}_{\nu_{\alpha}+\bar{\nu}_{\alpha}}}{\mathrm{d}E_\nu \mathrm{d}\Omega}\right] \,.
\end{equation}
In the following, we use this quantity to examine the potential detection and the projected constraints of heavy dark matter signals in upcoming neutrino radio telescopes.

\section{Next-generation neutrino telescopes}
\label{sec:neutel}

In the next decade many neutrino radio telescopes will be ready to start their data-taking phases, leading to a promising decade where some of the unknown properties of astrophysical UHE neutrinos could be unveiled. The aim of these neutrino radio telescopes is to detect the hadronic or electromagnetic showers that neutrinos create when interacting with a medium. This shower looks like a moving charge in a dielectric medium, leading to strong emission at radio wavelengths. This phenomenon is known as the Askaryan effect~\cite{Askaryan:1962hbi}. Due to the low neutrino fluxes expected at these ultra-high energies, large volumes of material are necessary for an efficient detection. For this reason the main media that have been used are air, ice or water. Among the promising future experiments which will be devoted mainly to neutrino observations, we focus on some benchmark ones, namely the Radio Neutrino Observatory in Greenland (RNO-G), the radio array of IceCube-Gen2, and the Giant Radio Array for Neutrino Detection (GRAND).

RNO-G~\cite{Aguilar:2020xnc} will be an ice-based neutrino telescope consisting of 35 stations with tens of radio antennas each. They aim at measuring the radio emission of the in-ice generated showers induced by UHE neutrinos through the Askaryan effect. Both electromagnetic and hadronic showers produce radio emission making the detector sensitive to all flavours, with roughly equal effective volumes at ultra-high energies for the radio detection method.

The same detection mechanism will be employed by IceCube-Gen2 radio array~\cite{Aartsen:2019swn, Aartsen:2020fgd}, which will be the next generation IceCube neutrino observatory in Antartica. Its radio array, which will consist of 200 stations covering a total area of 500 km$^2$, is expected to improve the RNO-G sensitivity by almost an order of magnitude.

GRAND~\cite{Alvarez-Muniz:2018bhp} is a planned observatory of ultra-high energy cosmic rays, gamma rays and neutrinos. It will be sensitive to the radio emission from EAS produced by ultra-high energy particles. The most promising strategy of neutrino detection focuses on the Earth-skimming underground tau neutrinos. The latter are astrophysical tau neutrinos passing through the Earth at very small angle with the surface, or passing through the mountains. In propagating through the rocks, neutrinos can produce tau leptons \textit{via} charged current interactions. The tau leptons can then exit the rock and decay into the atmosphere producing an EAS which is the target of detection of GRAND. A subdominant channel of detection could be based on horizontal neutrinos passing through large lengths in the atmosphere. The GRAND Collaboration estimates that the rate of events from this channel will be about one-tenth of the underground events. While the final stage of GRAND is planned to be composed of 200000 antennas, which is referred to as GRAND200k, we also discuss in this paper the intermediate stage of construction GRAND10k which will be composed by 10000 antennas. 

In these neutrino radio telescopes, possible background contamination includes incoherent thermal noise, impulsive anthropogenic noise, and mainly ultra-high energy cosmic rays misreconstructed as neutrino events. In GRAND200k, the rate of background events is estimated to be less than 0.1 per year. For GRAND10k this can probably be conservatively scaled down by a factor of 20. The background in RNO-G and IceCube-Gen2 is instead expected to be at the level of less than 0.01 events per station. A high level of event purity will be also ensured by a surface detector for cosmic rays and by the deployment of radio antennas in the ice at a distance of about $100~\mathrm{m}$ from the surface. This positional strategy has been validated by ARA neutrino telescope, which has achieved a background of 0.01 events in two stations over 1100 days~\cite{Allison:2015eky}. Hence, motivated by such a low expected background contamination, we can reasonably neglect it in the present analysis.

\section{Dark matter detectability}
\label{sec:detect}

In this section we first investigate the potential detection of dark matter neutrino signals in the aforementioned neutrino radio telescopes under the assumption of negligible contamination from background and potential astrophysical sources. The latter will be introduced in the next section when discussing the projected constraints on dark matter properties. Pursuing a phenomenological approach, we assume that DM particles can decay into a pair of particles of the Standard Model $\mathrm{DM} \to f \bar{f}$ with a 100\% branching ratio. In particular, we consider three different decay channels, namely into $b$ quarks, $\tau$ leptons and directly into neutrinos. These different scenarios are representatives of hadronic-philic, leptophilic and neutrinophilic dark matter particles, respectively. For the $\nu$ channel, we take equipartition among neutrino flavours at production.
\begin{figure}
    \centering
    \includegraphics[width=\textwidth]{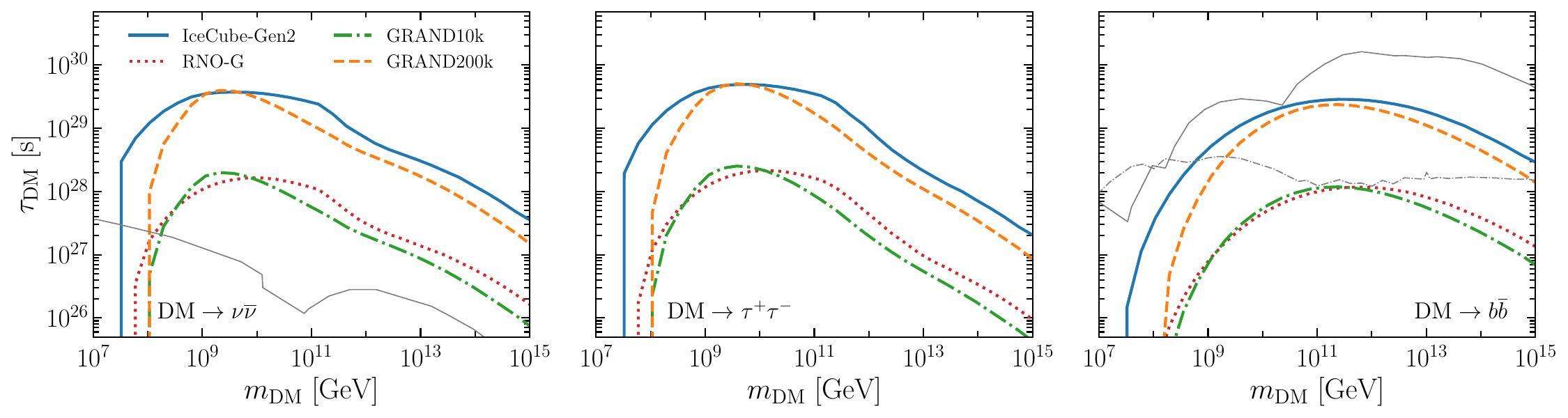}
    \caption{Potentiality for detection of a dark matter signal at the upcoming neutrino telescopes in 3 years of observation. The lines show the 95\% CL region for dark matter detection under the assumption of negligible contamination from background and astrophysical sources. Each panel corresponds to a different decay channel of DM particles. The thin gray lines are the existing constraints in the literature: a) for the $\nu\bar{\nu}$ channel with neutrino data from IceCube, PAO and ANITA~\cite{Esmaili:2012us}; b) for the $b\bar{b}$ channel with galactic multimessenger data~\cite{Ishiwata:2019aet}; c) for the $b\bar{b}$ channel with extragalactic multimessenger data~\cite{Ishiwata:2019aet}.
    \label{fig:dmdetectability}}
\end{figure}

For each dark matter model considered, the expected number of neutrino events in a given telescope after an exposure time $T_\mathrm{obs}$ can be computed as
\begin{equation} \label{eq:dmevents}
    n_\mathrm{DM}\left(m_\text{DM},\tau_\text{DM}\right) = T_\mathrm{obs}\int \mathrm{d}E_\nu \,\frac{\mathrm{d}\Phi^\text{DM}_{3\nu}}{\mathrm{d}E_\nu} A_{\rm{eff}} (E_\nu) \,.
\end{equation}
where the dark matter neutrino flux is defined in Eq.~\eqref{eq:DMneuflux} and $A_{\rm{eff}}$ is the effective area of the experiment averaged over neutrino flavours. If not explicitly reported by the experimental collaboration, we obtain an estimate of the detector effective area from the flux sensitivity~\cite{Alvarez-Muniz:2018bhp,Aartsen:2019swn,Aguilar:2020xnc}. In particular, being $S(E_\nu)$ the flux sensitivity, we take
\begin{equation}
    A_{\rm{eff}}(E_\nu) = \frac{2.44 E_\nu}{4\pi\ln \left(10 \right)\,S(E_\nu)\,T_\mathrm{obs}} \,,
\end{equation}
which provides 2.44 neutrino events for each energy decade. In the following analysis, we assume an exposure time of three years, namely $T_\mathrm{obs} = 3~\mathrm{yr}$. According to the Feldman-Cousins approach~\cite{Feldman:1997qc}, a detection of neutrinos from decaying dark matter at 95\% CL requires an expected number of neutrino events of at least 3.09. By equating Eq.~\eqref{eq:dmevents} to such a threshold value, we can obtain an upper value for the DM lifetime (for every DM mass) that every neutrino telescope can probe. 

In Fig.~\ref{fig:dmdetectability} we show the best sensitivity reach that can be achieved with 3 years of data-taking by the future neutrino telescopes RNO-G (dotted red lines), IceCube-Gen2 (solid blue lines), GRAND10k (dot-dashed green lines), and GRAND200k (dashed orange lines). The regions below the lines correspond to the dark matter parameter space that can be probed at 95\% CL under the assumption of negligible contribution from other sources. The three panels correspond to the different decay channels considered. If present, the gray lines display the existing constraints deduced by present neutrino telescopes as well as by gamma-ray and cosmic-ray observations. In particular, for the neutrino channel ($\mathrm{DM} \to \nu \bar{\nu}$), limits on the DM lifetime have been placed by analyzing the data taken by IceCube, PAO and ANITA experiments~\cite{Esmaili:2012us} (solid gray line in the leftmost plot). A more recent study based on updated neutrino and gamma-ray data is provided by ref.~\cite{Kachelriess:2018rty}. However, we note that their $\gamma$, $e^\pm$ and $\nu$ spectra from DM decays do not agree with the ones used in the present analysis. In particular, they are much harder at lower energies with respect to the ones computed with the \texttt{HDMSpectra} package~\cite{Bauer:2020jay}, and consequently lead to stronger limits on DM lifetime. Since a consistent comparison with our results is not feasible, in order to avoid confusion, we decided to not show their constaints in the plots. A deeper investigation of the existing inconsistencies among different computations of dark matter spectra is left for future work. For the decay channel into bottom quarks (rightmost plot), the dashed (b) and dot-dashed (c) lines correspond to the constraints placed by a recent very-detailed analysis of existing galactic and extragalactic multimessenger data, respectively~\cite{Ishiwata:2019aet}. For the tau channel we have not found similar multimessenger constraints in the literature. Hence, we find that, in the case of leptophilic and neutrinophilic scenarios, upcoming neutrino radio telescopes will be able to probe a still unexplored parameter space of heavy decaying dark matter. On the other hand, the detectable signals from hadronic-philic DM particles are already disfavored mainly by gamma-ray observations. However, it is worth mentioning that galactic dark matter searches typically suffer from large systematic uncertainties due to our little understanding of the dark matter halo in the inner regions of the Milky Way~\cite{Benito:2019ngh,Benito:2020lgu}. On the other hand, our results are expected to be robust to these uncertainties since they are based on the angle-integrated dark matter neutrino flux.

\section{Dark matter constraints}
\label{sec:ConsCons}

\begin{figure}
    \centering
    \includegraphics[width=0.75\textwidth]{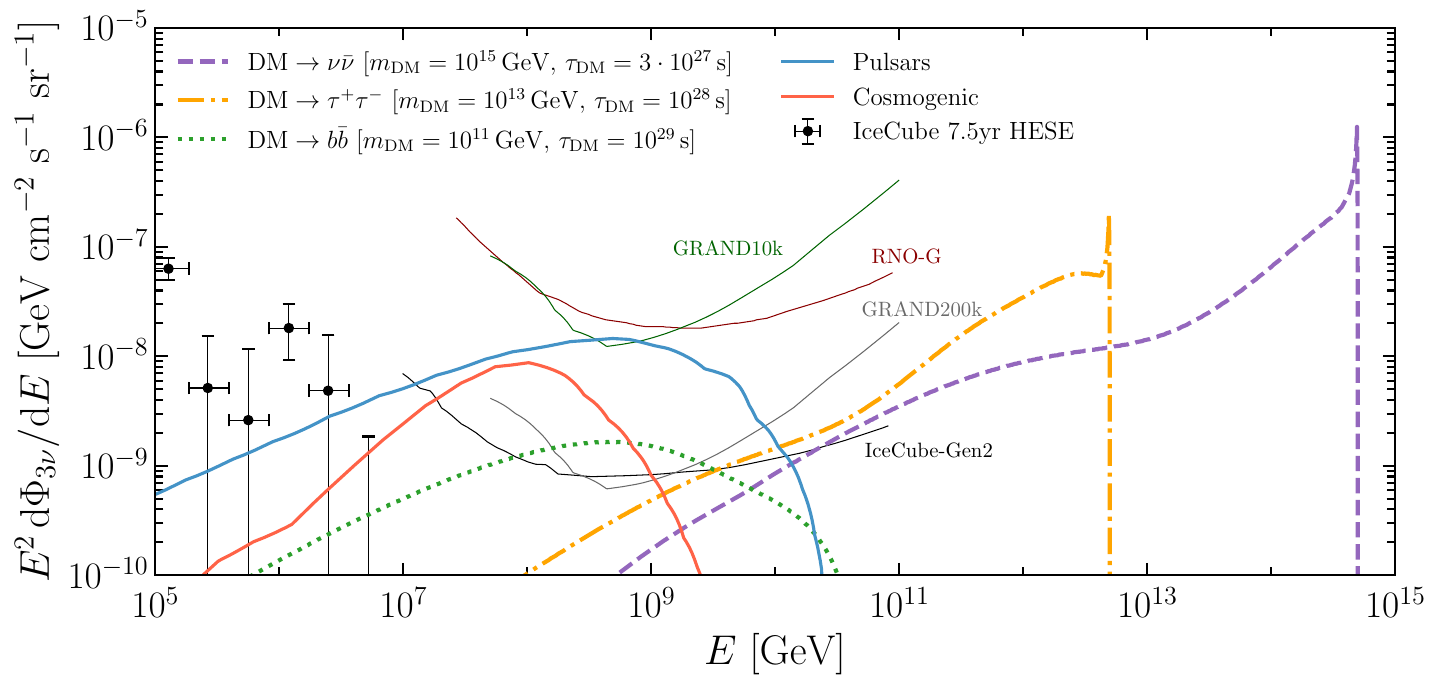}
    \caption{All-flavour neutrino fluxes predicted in two astrophysical scenarios (red line for cosmogenic neutrinos and blue line for newborn pulsar neutrinos) and in three different benchmark cases of decaying dark matter. The thin lines show the 3-year sensitivity reach of upcoming neutrino radio telescopes, while the data points correspond to the 7.5~year IceCube HESE sample~\cite{Abbasi:2020jmh}.}\label{fig:fluxes}
\end{figure}

In this section we describe the methodology adopted to estimate the sensitivity of future neutrino telescopes to constrain the parameter space of decaying dark matter. This information is complementary to the detection prospects discussed in the previous section. In order to place lower bounds on the lifetime of heavy DM particles, we here assume that future neutrino radio telescopes will observe UHE neutrinos coming from astrophysical sources. We rely on different theoretical models to predict the distribution of the expected number of events from astrophysical sources. In particular, we consider two different scenarios:
\begin{itemize}
    \item {\bf Cosmogenic neutrinos.} They constitute a guaranteed contribution to the UHE neutrino flux originated by the collision of ultra-high-energy cosmic rays on the CMB. However, the magnitude of this contribution is subject to severe theoretical uncertainties. For this work, in order to be conservative, we use the largest estimate shown in Ref.~\cite{Alvarez-Muniz:2018bhp} for the cosmogenic flux. As will be more clear later, a larger flux from astrophysical sources would result in weaker constraints on dark matter.
    \item {\bf Newborn pulsar neutrinos.} In addition to cosmogenic neutrinos, UHE neutrinos could also be produced in extreme astrophysical environments. Recent studies on astrophysical sources of UHE neutrinos include active galactic nuclei~\cite{Murase:2014foa,Rodrigues:2020pli}, gamma-ray bursts~\cite{Murase:2007yt}, flat-spectrum radio quasars~\cite{Righi:2020ufi}, black-hole jets embedded in large-scale structures~\cite{Fang:2017zjf} and newborn pulsars~\cite{Fang:2013vla}. In this work we also consider the neutrino component from newborn pulsars as reported in Ref.~\cite{Alvarez-Muniz:2018bhp}, which gives rise to the highest predicted flux and, consequently, leads to more conservative constraints on dark matter.
\end{itemize}

Figure~\ref{fig:fluxes} shows the all-flavour neutrino fluxes predicted by the two astrophysical scenarios, along with three benchmark fluxes from DM decay that could potentially be detected according to the results of the previous section. As a matter of comparison, we also show the sensitivity of the four neutrino radio telescopes under study. In the following, we first delineate in section~\ref{sec:method} the statistical approach adopted, and then discuss in section~\ref{sec:result} the constraints we obtain on the DM parameter space.

\subsection{Methodology} \label{sec:method}

\begin{figure}
    \centering
    \includegraphics[width=\textwidth]{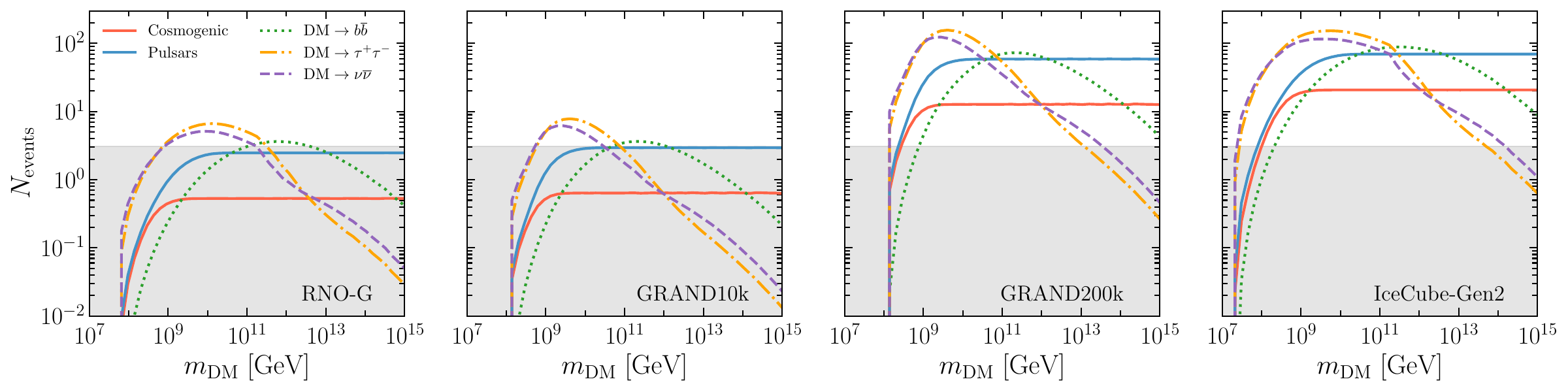}
    \caption{\label{fig:numevents} Expected number of neutrino events with an energy smaller than $E_\mathrm{max}=m_\mathrm{DM}/2$ in each neutrino telescope after 3 years of data-taking. The solid lines correspond to the astrophysical signal events while the dashed, dotted and dot-dashed are the ones corresponding to the expected signals coming from the different decay channels with a lifetime $\tau_\textrm{DM}=10^{28}~\mathrm{s}$. The number of astrophysical events is obtained by using Eq.~(\ref{eq:astroevents}) and fixing $E_\mathrm{max}=m_\mathrm{DM}/2$. The gray shaded area highlights the threshold of 3.09 events required for a detection at 95\% CL.}
\end{figure}

For each of the two scenarios (cosmogenic and newborn pulsars) discussed above, we need to compute the probability to observe a number $N_\mathrm{obs}$ of neutrino events in the upcoming experiments. Such a quantity is a stochastic variable which we assume to follow a Poisson distribution with mean $N_\mathrm{astro}$. The probability to observe $N_\mathrm{obs}$ events is therefore given by
\begin{equation}\label{eq:poisson}
    p\left(N_\mathrm{obs}|N_\mathrm{astro}\right)=\frac{\left(N_\mathrm{astro}\right)^{N_\mathrm{obs}} e^{-N_\mathrm{astro}}}{N_\mathrm{obs}!} \,.
\end{equation}
Similarly to Eq.~\eqref{eq:dmevents}, the mean value $N_\mathrm{astro}$ can be computed as
\begin{equation}\label{eq:astroevents}
    N_\mathrm{astro}\left(E_\mathrm{max}\right) = T_\mathrm{obs} \int_0^{E_\mathrm{max}}\mathrm{d} E_\nu\, \frac{\mathrm{d}\Phi^{\rm{astro}}_{3\nu}}{\mathrm{d}E_\nu} A_{\rm{eff}} (E_\nu) \,,
\end{equation}
where $\mathrm{d}\Phi^{\rm{astro}}_{3\nu}/\mathrm{d}E_\nu$ is the astrophysical all-flavour neutrino flux for each scenario. In order to increase the constraining power of our analysis, we only consider neutrinos with energy smaller than a maximum energy of $E_\mathrm{max}= m_\mathrm{DM}/2$. Indeed, $E_\nu \leq m_\text{DM}/2$ is the energy range in which we expect neutrinos from dark matter decays. Thus, restricting the analysis to this interval allows us to test the full DM signal while minimizing the number of expected astrophysical events. In Fig.~\ref{fig:numevents}, we report the mean values for the number of neutrino events from cosmogenic neutrinos (solid red lines) and newborn pulsars (solid blue lines) below a maximal energy $E_\mathrm{max}= m_\mathrm{DM}/2$ as a function of $m_\mathrm{DM}$, after three years of observations ($T_\mathrm{obs} = 3~\mathrm{yr}$). According to Eq.~\eqref{eq:astroevents}, the expected number of events from astrophysical sources increases as the maximum energy increases, until it reaches a constant value. This is due to the fact that both the astrophysical fluxes have an intrinsic cut-off energy above which no additional events are expected to be observed in the neutrino detectors. As can be seen, the newborn pulsar case produces a higher number of astrophysical neutrinos than the cosmogenic case. In the figure, we also show the expected number of events from dark matter particles with a benchmark lifetime $\tau_\mathrm{DM} = 10^{28}~\mathrm{s}$, for the three different channels. Moreover, the gray shaded area displays the threshold of 3.09 events required for detection at 95\% CL. Signals above such gray regions are therefore within the reach of the upcoming neutrino telescopes.

Given an observed number of events $N_\mathrm{obs}$, for each value of the DM mass we can compute the minimum lifetime such that neutrino signals from dark matter do not overshoot the expected astrophysical observation at 95$\%$ CL. In particular, we follow a procedure suggested in Ref. \cite{Cowan:2010js} that takes into account upper-fluctuations only. In order to be as conservative as possible, we constrain dark matter signals which predict a number of events $n(m_\text{DM},\tau_\text{DM})$ larger than the observed one. On the other hand, if $n(m_\text{DM},\tau_\text{DM}) < N_\mathrm{obs}$, we implicitly assume the existence of some unknown additional component that accounts for the observations. Our statistical approach is very conservative since any assumption on such an additional component would potentially lead to stronger constraints. We emphasize that, while the expected distribution of the observed data depends on the specific assumption for the astrophysical source of the observed neutrinos (see Eq.~\eqref{eq:poisson}), the determination of the constraints for a given observed number of events does not depend on it. In agreement with the above discussion, we consider the following test statistic (TS):
\begin{equation}
    \mathrm{TS}(m_\text{DM},\tau_\text{DM})=
    \begin{cases}~0 & \text{for} \quad n_\mathrm{DM} < N_\mathrm{obs} \\
    -2\ln\left(\frac{\mathcal{L}(N_\mathrm{obs} | n_\mathrm{DM})}{\mathcal{L}(N_\mathrm{obs}|N_\mathrm{obs})}\right)& \text{for} \quad  n_\mathrm{DM}\geq N_\mathrm{obs} \end{cases}
\end{equation}
where the likelihood $\mathcal{L}$ is assumed to be a Poisson distribution.

In order to reject the DM hypothesis, we compare the value of this TS with the probability distribution expected under the assumption that all the events were originated from decaying dark matter. This distribution is a half chi-squared distribution with one degree of freedom~\cite{Cowan:2010js}. With this information we can determine the value of the test statistic with which we can exclude the DM hypothesis at 95$\%$ CL, namely $\mathrm{TS}(m_\text{DM},\tau_\text{DM})=2.71$. Hence, for a fixed value of the DM mass and for a fixed value of the observed number of events $N_\mathrm{obs}$, we can determine the lower limit for the DM lifetime $\bar{\tau}_\mathrm{DM}\left(m_\text{DM}| N_\mathrm{obs}\right)$ below which a DM neutrino signal is excluded at $95\%$ CL. Since the number of observed events is a stochastic variable distributed according to Eq.~\eqref{eq:poisson}, the quantity $\bar{\tau}_\mathrm{DM}$ is also a stochastic variable. Its distribution is computed by means of the Poisson probabilities $p\left(N_\mathrm{obs} | N_\mathrm{astro}\right)$ for the two astrophysical scenarios considered. In this way, we can therefore determine the confidence intervals in which we expect the lower limit for the DM lifetime to lie. We adopt the conservative procedure of analyzing only the total number of events over the entire energy range $E_\nu \leq m_\mathrm{DM}/2$. Taking into account an energy and/or angular binning would allow one to also test the shape of the DM signal, and is generally expected to lead to stronger constraints. Moreover, examining the integrated neutrino spectrum makes our analysis more robust, being less sensitive to the experimental details of the upcoming neutrino telescopes.

\subsection{Results} \label{sec:result}

\begin{figure}
    \centering
    \includegraphics[width=\textwidth]{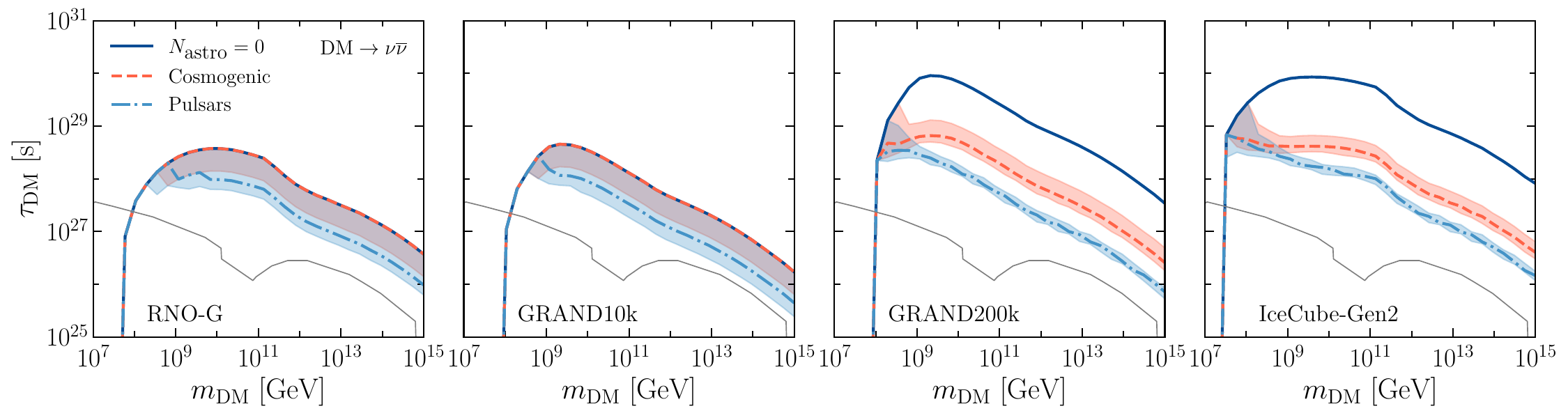}
     \vskip3mm
    \includegraphics[width=\textwidth]{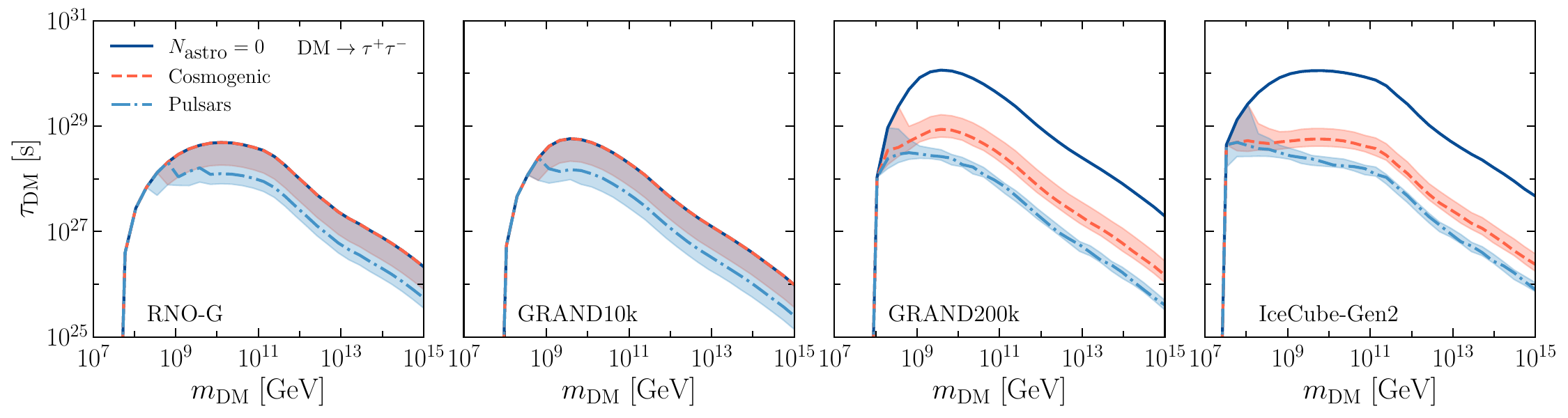}
    \vskip3mm
    \includegraphics[width=\textwidth]{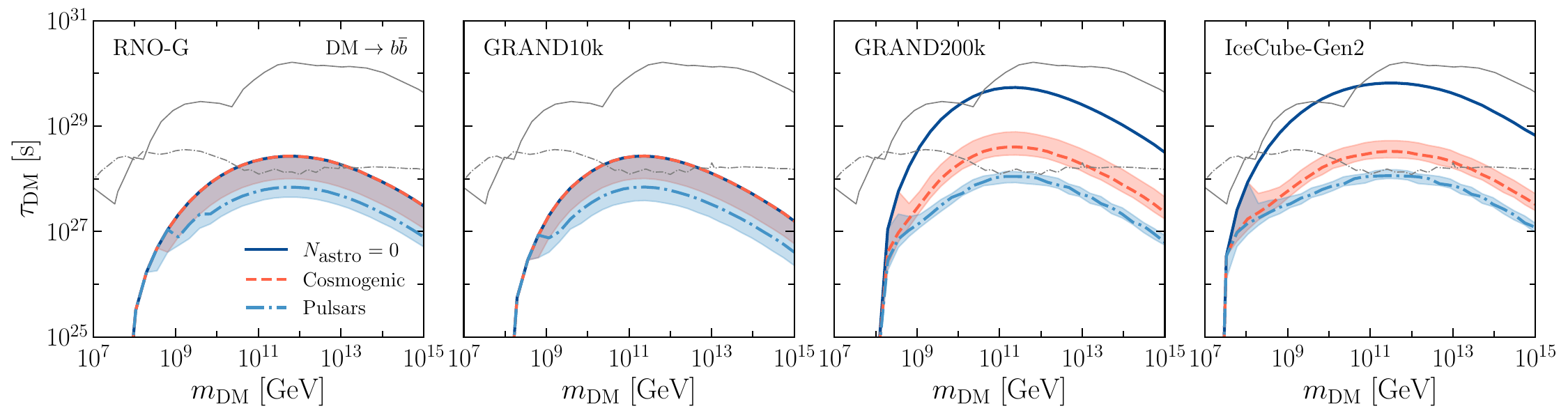}
    \caption{\label{fig:3yr} Projected 3-year constraints at 95\% CL in the plane $m_\mathrm{DM}$-$\tau_\mathrm{DM}$. Each row corresponds to a different DM decay channel, while each column to a different upcoming neutrino radio telescope. The bands represent the $2\sigma$ intervals according to the Poisson distribution in Eq.~\eqref{eq:poisson} of observing a given number of events in the cosmogenic (red color) and the newborn pulsars (blue color) scenario. The red dashed and blue dot-dashed lines show the most probable constraints on $\tau_\mathrm{DM}$ assuming the observation of astrophysical neutrinos. The dark-blue solid lines display the limits obtained in the case of zero detected events ($N_\mathrm{astro} = 0$). The thin gray lines are the existing constraints in the literature: a) for the $\nu\bar{\nu}$ channel with neutrino data from IceCube, PAO and ANITA~\cite{Esmaili:2012us}; b) for the $b\bar{b}$ channel with galactic multimessenger data~\cite{Ishiwata:2019aet}; c) for the $b\bar{b}$ channel with extragalactic multimessenger data~\cite{Ishiwata:2019aet}.}
\end{figure}

The main results of the present work are collected in Fig.~\ref{fig:3yr}, where we show the 2$\sigma$ bands of the lower limit for the DM lifetime at 95\% CL, for all the combinations of DM decay channel, experiment and astrophysical scenario adopted to simulate data. The dashed and dot-dashed colored lines represent the most probable lower limit according to the Poisson distribution in Eq.~\eqref{eq:poisson} under the assumption of observing cosmogenic and newborn pulsar neutrinos, respectively. Moreover, the dark-blue solid lines show the constraints that could be potentially placed in the case of zero detected events ($N_\mathrm{astro} = 0$). Such a case is realistic given the present knowledge: indeed the uncertainty band on the cosmogenic neutrino fluxes is so large that these fluxes could be below the sensitivity of the upcoming neutrino radio telescopes. The thin gray lines (if present) instead correspond to the existing constraints as reported in Fig.~\ref{fig:dmdetectability}. We find that the cosmogenic astrophysical signal case (red color) always leads to stronger constraints than the newborn pulsar case (blue color). This stems from the lower expected number of events in the cosmogenic case as shown in Fig.~\ref{fig:numevents}. This result is very pronounced in the case of large neutrino telescopes such as GRAND200k and IceCube-Gen2 which are able to distinctly detect the two astrophysical fluxes. On the other hand, in the case of RNO-G and GRAND10k with smaller effective volumes, the two bands (cosmogenic and newborn pulsar scenarios) overlap due to the low expected values for $N_\mathrm{obs}$.

In all the cases, the sharp cut-off occurring in the low DM mass range (especially for $\nu\bar{\nu}$ and $\tau^+\tau^-$ channels) is instead due to the very low sensitivity of the upcoming neutrino radio telescopes for $E_\nu \lesssim 10^{8}~\mathrm{GeV}$. At low DM masses, the $2\sigma$ bands reduce to a single line. This stems from the fact that the energy range considered to compute the number of observed events reduces as the DM mass decreases. Hence, since the bands are computed using the discrete Poisson distribution~\eqref{eq:poisson}, at low DM masses we expect $N_\mathrm{obs} = 0$ with a very high probability while $N_\mathrm{obs} = 1$ is very unlikely to occur. The discrete nature of the Poisson distribution also explains the abrupt discontinuity in the low DM mass range in the case of GRAND200k and IceCube-Gen2. Indeed, with the increase of the DM mass which enlarges the energy range of the analysis ($E_\mathrm{max} = m_\mathrm{DM}/2$), the probability of $N_\mathrm{obs}=0$ becomes smaller and smaller until it lies outside the 2$\sigma$ interval of the Poisson distribution. When this occurs, the limits abruptly weaken and move away from the dark-blue lines corresponding to $N_\mathrm{astro}=0$.

Before concluding this section, a comment concerning the possible presence of a background is in order. As discussed in section~\ref{sec:neutel}, the background contamination is expected to be very low in all the experiments considered. This motivates our choice of not including it in our forecast analysis. An additional background component would have the effect of increasing the number of expected neutrino events and consequently of weakening the lower limits in Fig.~\ref{fig:3yr}. With the current background estimates, we find that the constraints on the DM lifetime could roughly weaken by $\sim 30\%$ and $\sim 10\%$ for the cosmogenic and newborn pulsars case, respectively. Hence, the impact of the background contamination is by far within the 2$\sigma$ bands reported in Fig.~\ref{fig:3yr}, which makes our results very robust. Moreover, we note that the background contamination could be further reduced in the future thanks to a better understanding of the detectors performances and of the background energy and angular distributions.

\section{Conclusions}\label{sec:conclusions}

In the next decade we will witness the construction of neutrino radio telescopes capable of detecting neutrinos of unprecedented high energy, opening a window into the ultra-high-energy neutrino astronomy. While the main goal of these next-generation telescopes is the observation of cosmogenic neutrinos, they also have the potential to detect a contribution coming from decaying dark matter particles. In this work, we have investigated the future prospects of upcoming neutrino radio telescopes in terms of the expected sensitivity for decaying dark matter. In particular, we have focused on four benchmark telescopes: RNO-G, GRAND10k, GRAND200k and IceCube-Gen2 radio array. By assuming the observation of cosmogenic and newborn pulsar neutrinos (the highest neutrino fluxes proposed in the literature), we have performed a forecast statistical analysis in order to set conservative bounds on the lifetime of heavy dark matter particles with masses in the range of $10^7-10^{15}~\mathrm{GeV}$. For the sake of concreteness, we have analyzed three different decay channels which are representative of hadronic-philic, leptophilic and neutrinophilic dark matter particles.

By presenting our projections in the context of multimessenger analyses, we have demonstrated that upcoming neutrino telescopes will be able to probe new regions of the dark matter parameter space. Our results are robust and conservative since they have been obtained analyzing the total neutrino events without considering their angular and energy distribution. For the $b\bar{b}$ channel, most of the parameter space within reach of upcoming neutrino radio telescopes is already disfavored by cosmic-ray and gamma-ray data. On the other hand, we have found that, in case of leptophilic and neutrinophilic dark matter particles, the planned EeV neutrino telescopes are expected to improve the existing limits by a few orders of magnitude. It is worth emphasizing that indirect dark matter searches with neutrinos are highly complementary to the ones with the other messengers due to different theoretical and experimental uncertainties.

\section*{Acknowledgements}
This work was partially supported by the research grant number 2017W4HA7S ``NAT-NET: Neutrino and Astroparticle Theory Network'' under the program PRIN 2017 funded by the Italian Ministero dell'Universit\`a e della Ricerca (MUR). The authors also acknowledge the support by the research project TAsP (Theoretical Astroparticle Physics) funded by the Istituto Nazionale di Fisica Nucleare (INFN).

\bibliography{references}

\end{document}